\pdfoutput=1

\documentclass[11pt]{article}

\usepackage{acl}

\usepackage{times}
\usepackage{latexsym}

\usepackage[T1]{fontenc}

\usepackage[utf8]{inputenc}

\usepackage{microtype}

\usepackage{amsbsy}   
\usepackage{booktabs} 
\usepackage{caption}  
\usepackage{graphicx} 
\usepackage{amsfonts} 
\usepackage{color}
\usepackage{subfigure}
\usepackage{xurl}

\title{ Mental Health Self-Disclosure

on Social Media throughout the Pandemic Period

\vspace{.5em}

}

\author{Dino Husnic, ~ Stefan Cobeli, ~ Shweta Yadav  \\
University of Illinois at Chicago \\
  \texttt{\{dhusni2,scobel2,shwetay\}@uic.edu}
  }

\begin{document}
\maketitle
    \begin{abstract}
        
        The COVID-19 pandemic has created many problems, especially in people's social lives. 
        There has been increasing isolation and economic hardships since the beginning of the pandemic for people all over the world. 
        Quarantines and lockdowns also took part in that, and so, people have been expressing their emotions throughout the pandemic period using social media platforms like Reddit, Twitter, Facebook, etc. 
        In this study, we seek to analyze the emotions and mental health labels throughout the time period of March 2, 2020, up until July 4, 2020, from the threads and comments gathered from the r/unitedkingdom subreddit. 
        We used a soft labeling technique to generate mental health conditions for each Reddit comment.
        We compared the overall results with important dates related to COVID-19 policies that took place in the United Kingdom. 
        This can give us a view on how the pandemic and the important dates affect people self disclosing their emotions on social media platforms.
        Finally, we have developed a proof of concept to show that using mental health features may increase emotion prediction accuracy.
    \end{abstract}

    \section{Introduction}
    \label{sec:introduction}
    
        Certain emotions are related to certain mental health conditions. COVID-19 has gotten many people to express their emotions more online due to things like isolation or feelings of loneliness. Mental health could be recorded with emotions for important purposes, such as understanding what kinds of local/global events can impact the majority of people's mental states, and whether they are more positive or negative than before. They can also help to identify problematic posts as quickly as possible. Public announcements and events that are related to the pandemic could possibly affect the amount of mental health-related posts around the time of them, and could show how people’s emotions/mental health-related posting or self-disclosure can change. We can record these values and compare certain emotions with mental health, and see differences in both throughout different periods of time during the pandemic. This could also show us trends in overall posts over time, compared to an average week, such as most of the posts occurred during the beginning of the dataset's time period but there were spikes in posts in certain days/weeks since certain laws were enacted or global events relating to the pandemic occurred \cite{basile2021dramatic}. This could predict the overall emotional reactions to future announcements.
    
    In order to investigate the development of the mental health status of social media users, we considered a number of data sources. 
    First, we note the fact that there were multiple large-scale social media datasets crawled with posts spanning throughout the pandemic period.
    For example, \newcite{chen2020tracking} crawled a dataset with over $100$ million Twitter posts beginning from January $2020$ all the way to $2022$\footnote{The tweets' information is freely accessible on Github: \url{https://github.com/echen102/COVID-19-TweetIDs}}.

    Another large-scale dataset with over $1.12$ billion Twitter posts was developed by \newcite{banda2021large}.
    The authors filtered the Twitter posts related to COVID-19 and released the obtained results publicly available.
    Moreover, the dataset is updated with daily posts\footnote{The dataset is also available on Github \url{https://github.com/thepanacealab/covid19_twitter}}.

    In order to benchmark our analysis, we used the dataset provided by
    \newcite{basile2021dramatic}\footnote{The Reddit dataset can be accessed at \url{https://bitbucket.org/cauteruccio/reddit-dataset/}}.
    The dataset consisted of six different subreddits from March 2020 up until July 2020. The subreddits looked at were r/italy, r/de, r/sweden, r/unitedkingdom, r/thenetherlands, and r/nyc.
    Plutchik's Wheel of Emotions was looked at to determine the 8 emotions that were going to be used: joy, trust, fear, surprise, sadness, disgust, anger, and anticipation. Skepticism was also recorded as an opposite emotion to trust, but it was not included in the 8-dimensional array. Once they created a model using NRC-EIL (NRC Emotion Intensity Lexicon), they used a SemEval test dataset with around 8,000 tweets, 4,000 for testing, and got an F1 score of 0.42 \cite{basile2021dramatic}.
    We focused on the fragment of the dataset crawled within the United Kingdom. 
    This dataset fragment contained $111,317$ Reddit comments from  $6,229$ users.

    To investigate the potential relationship between mental health conditions and emotions, we used an adaptation of the \textsc{MentalRoBERTa} model \cite{ji2021mentalbert} for sequence classification\footnote{Models are available on the \textsc{huggingface} website:
    \url{mental/mental-roberta-base} and 
    \url{rabiaqayyum/autotrain-mental-health-analysis-752423172}}.
    We used this model to classify the dataset with seven different labels: depression, anxiety, BPD, bipolar, mentalhealth, autism, and schizophrenia. 
    There was a reported accuracy score of 80.52\% from this model.
    
     \begin{figure}[!ht]
        \centering
        \includegraphics[width=.45\textwidth]{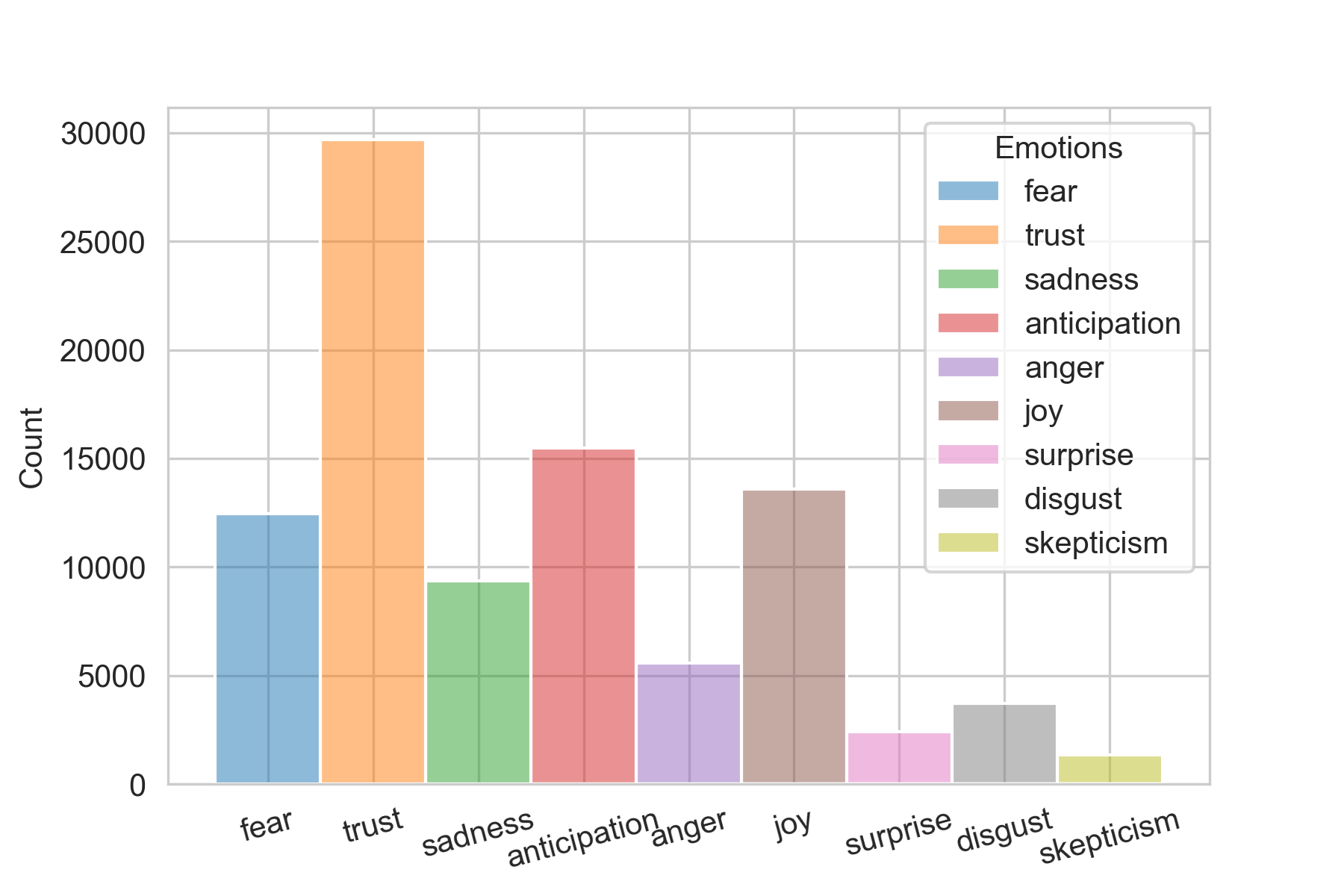}
        \caption[]{Emotions distribution in the Reddit dataset when considering the tweets crawled in the \texttt{/r/unitedkingdom} subreddit. 
        }
        \label{fig:emo_ditribution}
    \end{figure}

    \paragraph{Related Work}
    In recent years, social media has emerged as an active platform to study mental health problems \cite{mazhar2025figurative, yadav2023towards,yadav2020identifying}, substance use \cite{babaeianjelodar12023around,chen2022us,chen2022vaping,yadav2021they,lokala2021social}, and other healthcare issues \cite{agarwal2025overview,bahri2024consumer,yadav2020identifying,yadav2021reinforcement,yadav2022chq,yadav2023towards,yadav2022towards,zhang2022focus,zhao2024heterogeneous,naik2024no}. 
    Numerous studies examine the social effects of the pandemic. The analyses span from the dawn days, in December 2019, and the fear expressed on the WeChat social media platform, in China  
    \cite{allam2020first}, 
    to emergent themes in the US Twitter discussions
    \cite{valdez2020social} and
    all the way to far-reaching human existential questions expressed on a vast majority of media platforms 
    \cite{sathish2020report}.
    
    On the other hand, the pandemic also had an impact on lighter subjects, such as political opinions
    \cite{chum2021changes}
    and even on passionate travelers' expertise expressed on Reddit 
    \cite{hao2022social}.
    
    \section{Dataset Description}
    \label{sec:dataset}
    
    \begin{figure}[!h]
        \centering
        \includegraphics[width=.45\textwidth]{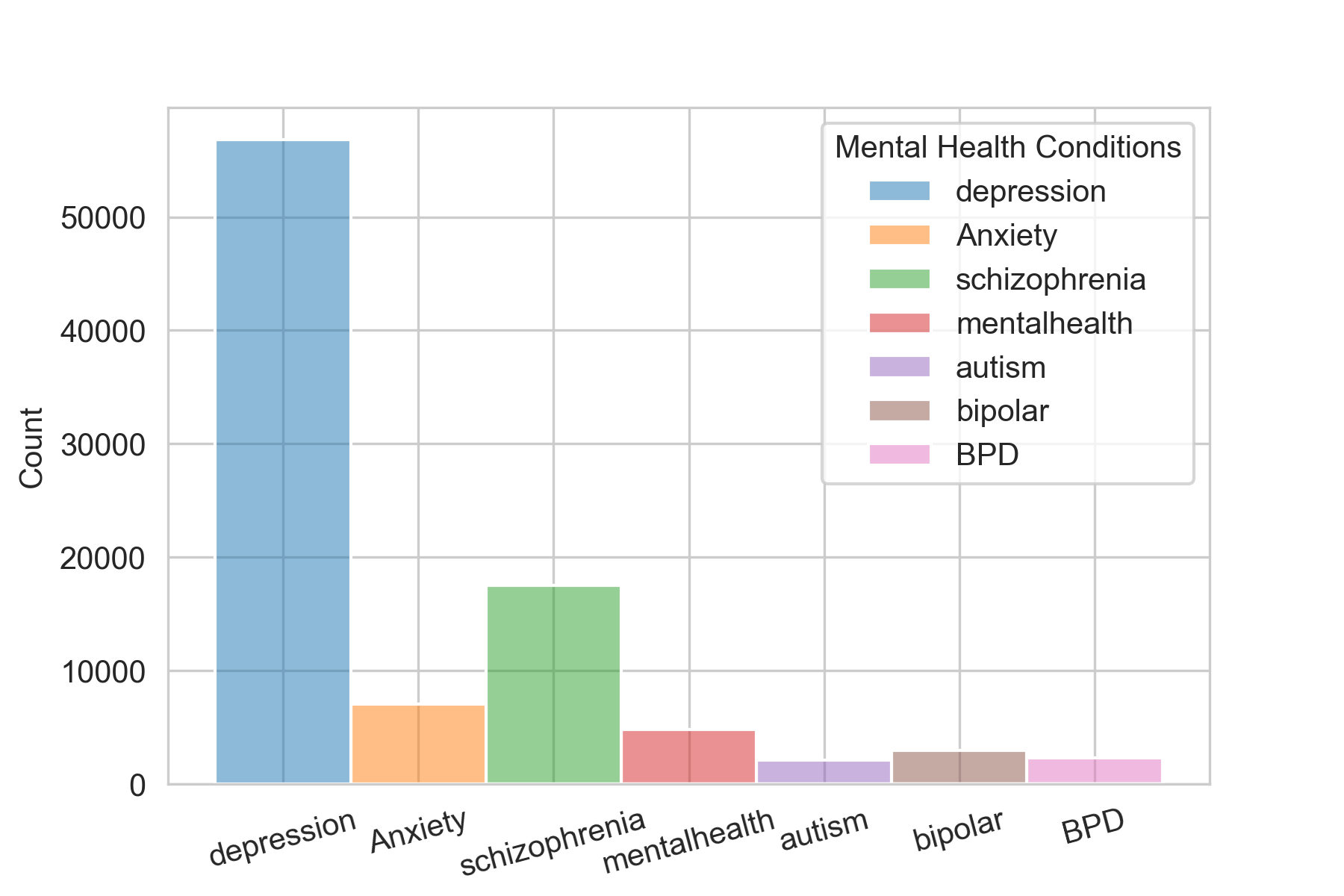}
        \caption[]{Mental health condition distribution in the \texttt{/r/unitedkingdom} subreddit after using our soft labeling technique.
        }
        \label{fig:mh_distribution}
    \end{figure}

    As we mentioned in \autoref{sec:introduction}, we used the Reddit dataset \cite{basile2021dramatic} with posts from the United Kingdom subreddit.
    The authors labeled the Reddit comments with one of $8$ possible emotions according to the \texttt{NRC} lexicon \cite{mohammad2017word}.  
    Concretely, each comment $c$ was labeled with $\sum_{t \in c} \texttt{NRC}(t)$ where $t$ are the token words in the comment $c$ and $NRC(t)$ is the eight-dimensional vector embedding of a token $t$ according to the \texttt{NRC} lexicon.
    Furthermore, the authors introduced a ninth emotion specific to the COVID-19 period, namely \textit{skepticism}.
    The skepticism score was computed similarly to the other emotions, but the lexicon for skepticism was defined as being formed out of all antonyms of the words associated with the \textit{trust} emotion.
    Thus, the authors obtained a nine-dimensional vector corresponding to anger,
    fear, disgust, sadness, surprise, trust, anticipation, joy, and skepticism for each Reddit comment in the dataset. 
    In \autoref{fig:emo_ditribution}, we displayed the distribution of the emotions in the Reddit dataset when considering the label of a tweet to be the maximum of the nine components of the emotion scores.

    \begin{figure}[!h]
        \centering
        \includegraphics[width=.45\textwidth]{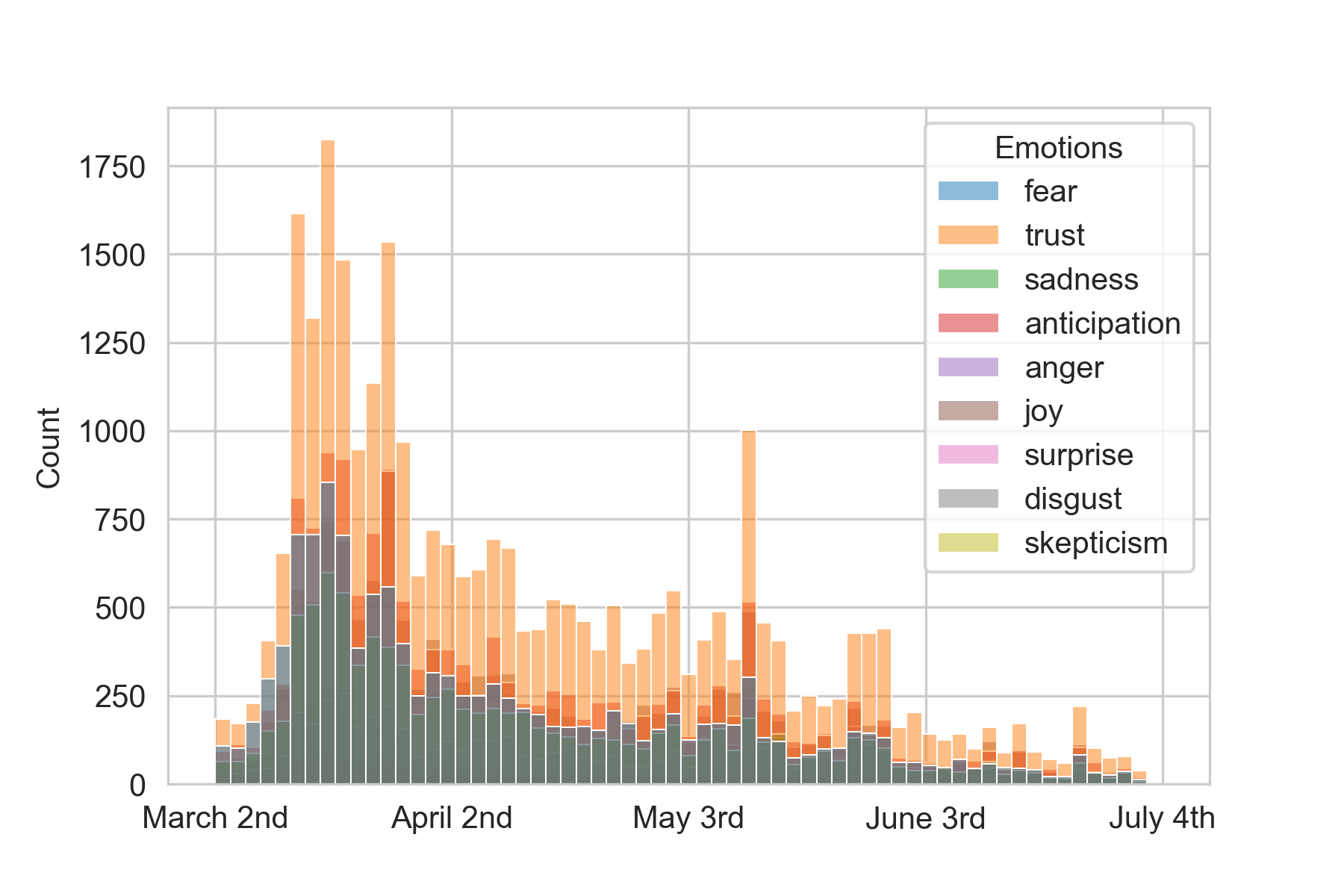}
        \caption[]{ Distribution of the expressed emotions in the 
        \texttt{/r/unitedkingdom} subreddit comments.}
        \label{fig:emotion_timeline}
    \end{figure}
    
    \subsection{Mental Health Labeling}
    
    Our work's goal was to analyze the development of the mental health conditions of social media users in social media posts.
    To infer the mental health labels of the posts in our dataset, we employed a soft label technique.
    We used a pretrained \textsc{Mental-RoBERTa} model for sequence classification to generate mental health labels for each Reddit post.
    The mental health labels we generated for each Reddit comment were selected from 7 possible conditions: \textit{Anxiety, BPD, Autism, Bipolar, Depression, Mental Health} and \textit{Schizophrenia}.
    We computed the softmax of the \textsc{Mental-RoBERTa} computed logits, and we labeled each Reddit comment with the condition having the maximum softmax score.
    By doing so, we obtained the mental health conditions distribution displayed in \autoref{fig:mh_distribution}.
 We can notice in \autoref{fig:mh_distribution} that most Reddit comments were labeled as depressive.
    Still, it is important to consider that our soft labeling technique does not take into account posts that might not display any mental health condition.
    In future studies, we will need to mitigate this shortage by at least adding another label representing a general post which does not express any mental health disorder.
    
    \section{Timeline Analysis}
    
    In this section, we will display our analytical results representing the development of mental health conditions in the Reddit posts throughout the pandemic period.
    
    
    \begin{figure}[!h]
        \centering
        \includegraphics[width=.45\textwidth]{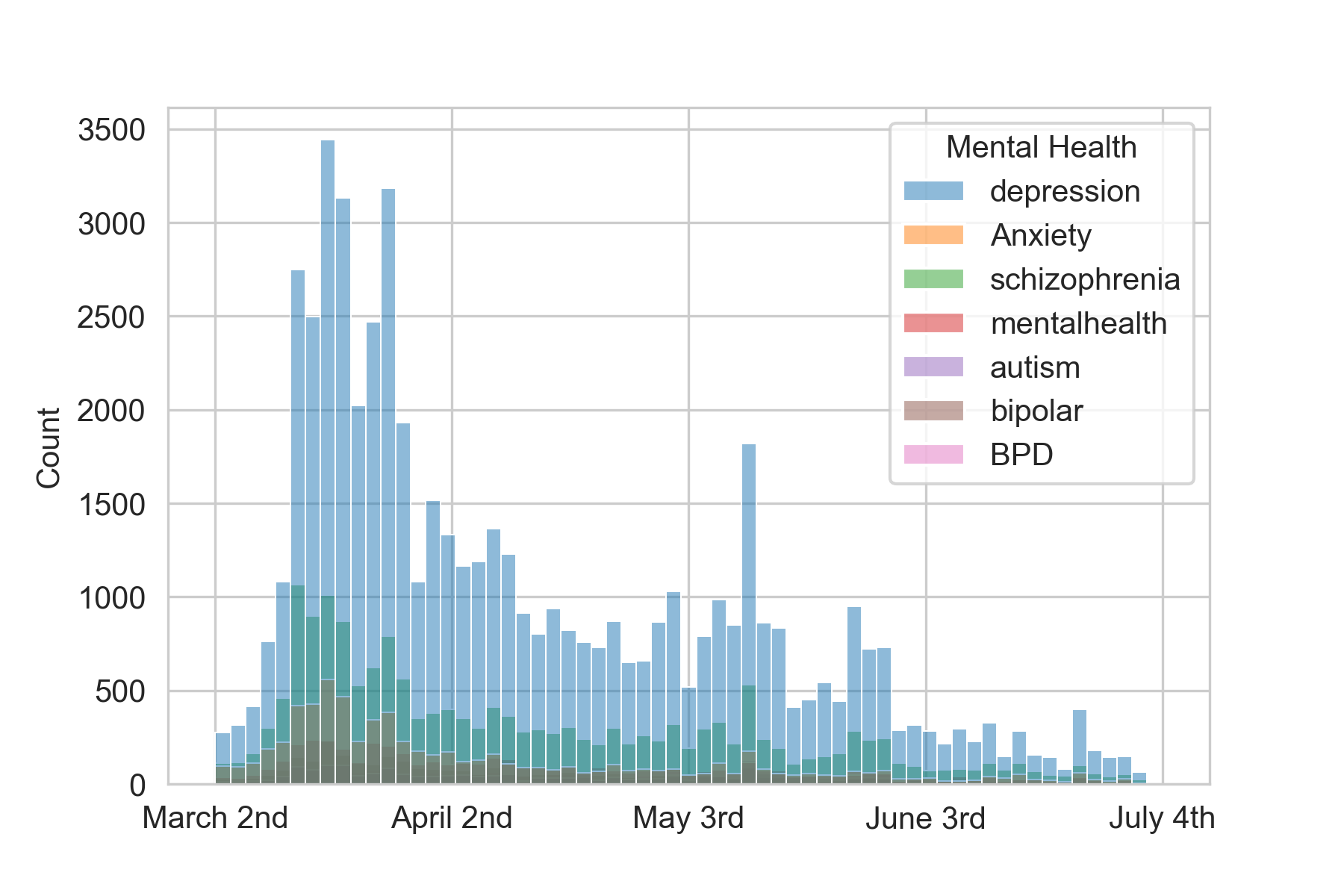}
        \caption[]{Distribution of the expressed mental health conditions in the 
        \texttt{/r/unitedkingdom} subreddit comments.
        }
        \label{fig:mh_timeline}
    \end{figure}
    
    

    \subsection{Entire Timeline}
    
    We begin by first discussing the distribution of emotions and mental health conditions throughout the entire period covered by the dataset.
    
    First, we can see in \autoref{fig:emotion_timeline} the frequency of Reddit comments labeled with one of the nine available emotions. 
    We can observe that there is no significant difference between the proportions of the expressed emotions on each day.
    We can interpret the mode of the displayed distribution by noticing it occurred when the lockdown was initiated, and people started staying home for longer periods.
    As the restrictions were dropped, the number of Reddit posts also decreased.
    
    The main days that will be looked at are March 23, 2020 (the day the lockdown was set), April 22, 2020 (the first restrictions lifted), and June 15, 2020 (shops open).

    Similarly to the frequency of emotions, we displayed the distribution of mental health conditions expressed in Reddit comments in \autoref{fig:mh_timeline}.
    Again, we can see that there is no significant difference between the daily proportions of the expressed mental health conditions.

     \begin{figure*}[!ht]
        \centering
        \includegraphics[width=.95\textwidth]{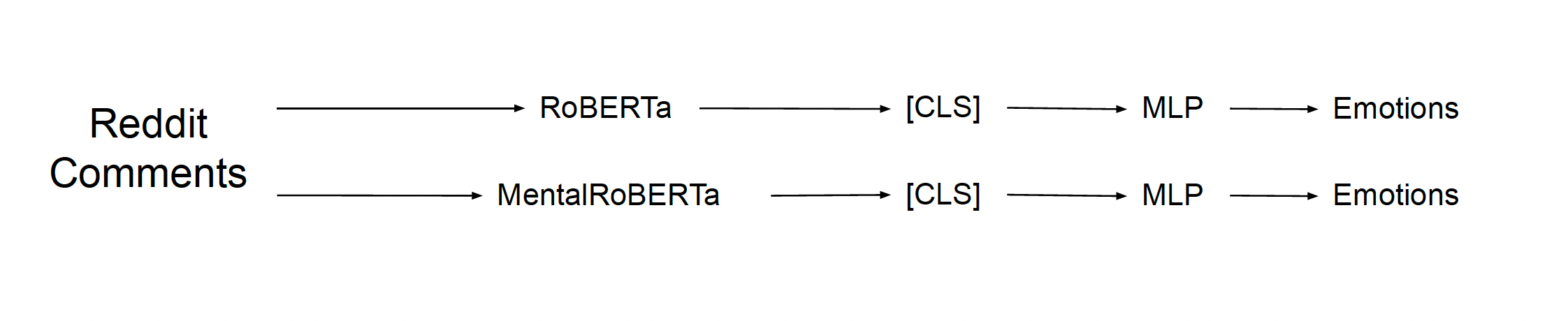}
        \caption[]{
        Proposed methodology architecture to predict emotions of Reddit comments using the pretrained embeddings of RoBERTa and MentalRoBERTa.
        }
        \label{fig:proposed_architecture}
    \end{figure*}
    
    
    The next analysis we performed was that of mental health conditions during the weeks of the pandemic.
    We have displayed the average occurrence of each mental health condition in a general week versus the average during a week characterized by a notable event. 
    We also noticed that during an usual week, anxiety levels tended to drop on Sunday.
    
    On the other hand, in the week when the lockdown was announced in the UK, the anxiety level increased at a significantly larger pace.
    Likewise, we observed that when some of the first restrictions were lifted, the agitation level increased significantly, and thus the schizophrenia levels spiked noticeably.
    We also analyzed a week from the end of the first pandemic wave.
    
\begin{figure}[!ht]
        \centering
        \includegraphics[width=.45\textwidth]{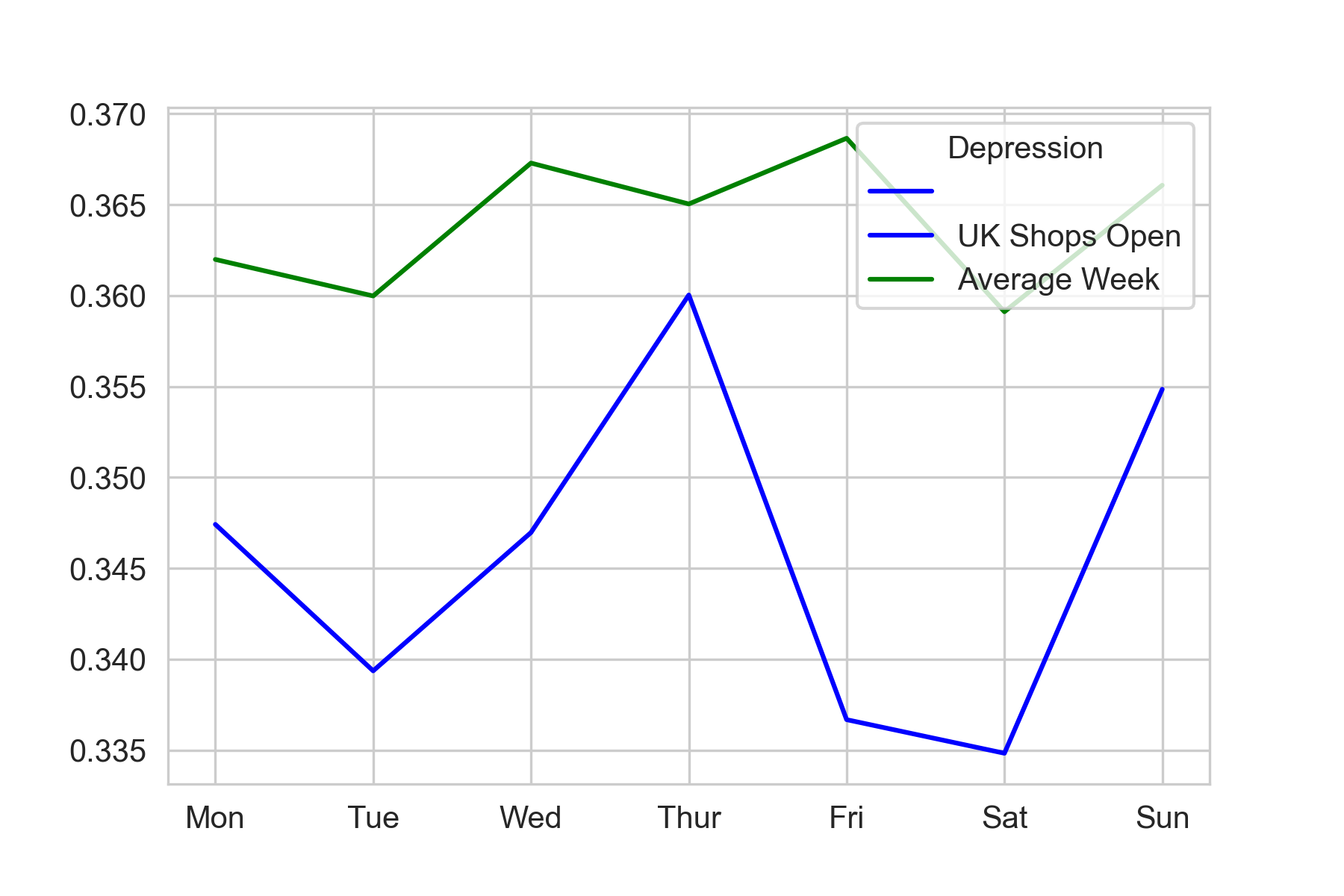}
        \caption[]{Comparison between the frequency of depression labeled comments during an average pandemic week (green) and the depression levels expressed in one of the final weeks of the first wave, when shops were open in the UK.
        }
        \label{fig:week_depression}
    \end{figure}

\begin{figure}[!ht]
        \centering
        \includegraphics[width=.45\textwidth]{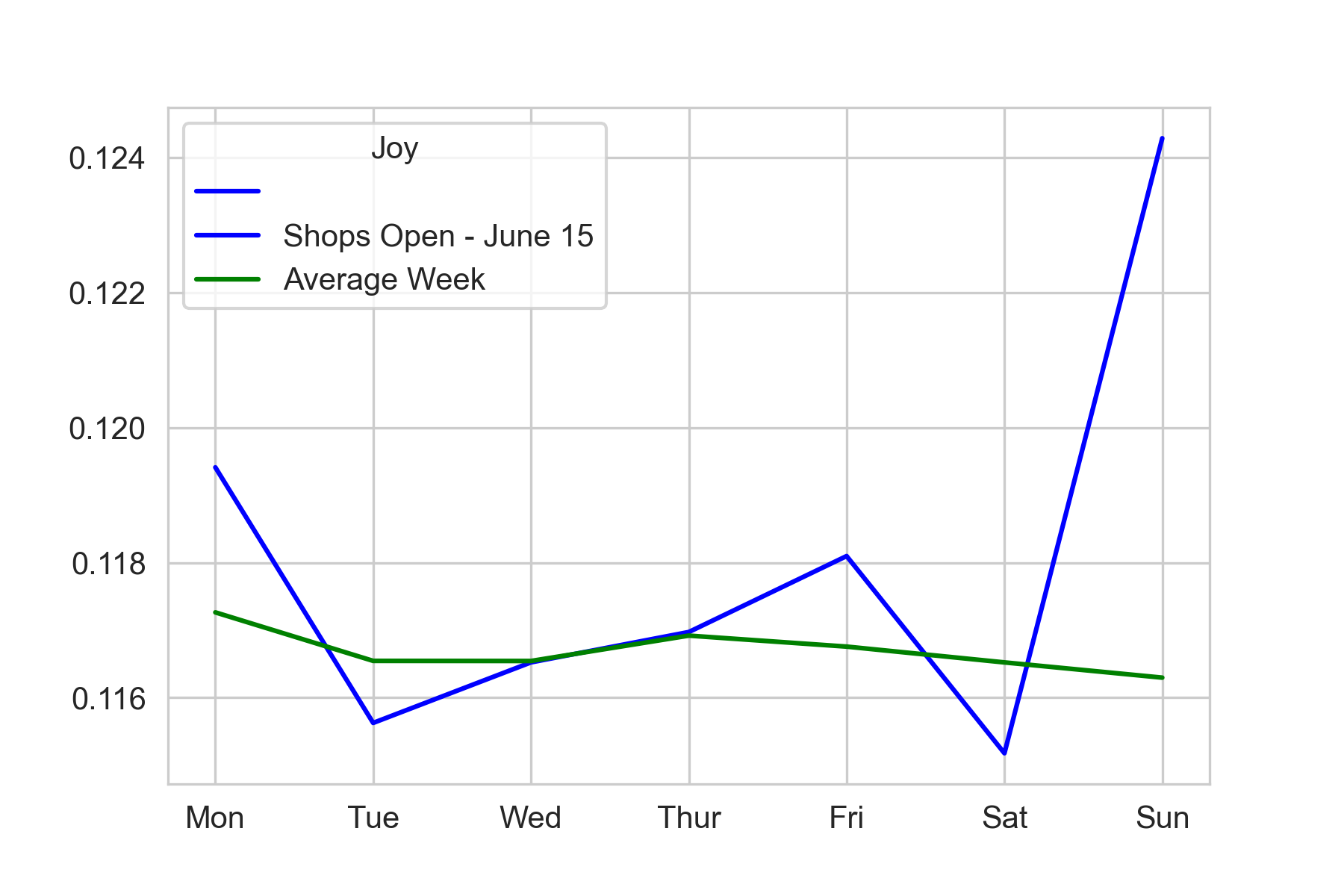}
        \caption[]{Comparison between the frequency of joy-labeled comments during an average pandemic week (green) and the joy levels expressed in one of the final weeks of the first wave, when shops were open in the UK.
        }
        \label{fig:week_joy}
    \end{figure}

    In \autoref{fig:week_depression}, we can see that the depression levels in the week when the UK shops were opened were at significantly lower levels than in a usual previous week. 
    Finally, in \autoref{fig:week_joy} we can see that joy levels spiked in the weekend when the shops were lifted, marking the end of the first COVID-19 wave in the UK.
    
    \section{Predicting Emotions using Mental Health Features}
    
    One of the analysis directions we were interested in for this study was the correlation between emotions and mental health conditions.
    Since the provided Reddit dataset was softly labeled with emotions, we had a valuable starting point.
    
    We proposed the following methodology.
    Given a Reddit comment $C$, we used the [CLS] embeddings of $C$ produced by RoBERTa $C_{CLS}^{R}$ and the [CLS] embeddings of $C$ produced by MentalRoBERTa $C_{CLS}^{MR}$.

    \begin{table}[!ht]
	  \centering
	  {\begin{tabular}{lcc}
	    \toprule
	    Model  &  F1 Score. & Accuracy.\\
	    \midrule
	    RoBERTa  CLS        & $31.31$ & $35.54$\\
	    MentalRoBERTa CLS     & $\pmb{32.37}$ & $\pmb{38.92}$ \\
	    \bottomrule
	  \end{tabular}}
	  \caption{
	    Performance of emotion label prediction of reddit comments using the CLS embedding of either RoBERTa or MentalRoBERTa.
	  }
	  \vspace{-3mm}
	  \label{table:results-BERT}
	\end{table}	
	
    \section{Conclusions and Future Work}
    
    In the current study, we have aimed to explore the trends in the mental health conditions expressed on social media using a readily available Reddit dataset.
    We used a pretrained MentalRoBERTa model to generate soft labels for Reddit comments.
    We also showed that training a model on top of the mental features is more beneficial for emotion prediction than using a vanilla pretrained language model.
    
    Finally, we wish to highlight a couple of insufficiencies in the current study and also propose ideas that remain to be developed.
    We are further interested in what topics determine specific mental health conditions.
    By using topic modeling, we could further get insights into the mental health conditions expressed in social media posts.
    
    There are many more subreddits that can be looked at, other than the ones in the dataset, in order to see mental health-related postings. 
    Specific mental health subreddits could be broken down into emotions as well, and compared to just the mental health disorder.
    The pandemic did not end in July of 2020, so there can be a much larger time frame that can be looked at, with even more important events, like the first vaccine approved or a specific variant increasing cases. 
    The approach used here could also be adapted in order to translate to other social media platforms and collect data from them.

\bibliography{anthology,custom}
\bibliographystyle{acl_natbib}



    

	





\end{document}